\documentclass[reprint,superscriptaddress,amsmath,amssymb,aps]{revtex4-1}
\usepackage{graphicx,textcomp}
\usepackage[unicode=true,pdfusetitle,bookmarks=true,bookmarksnumbered=false,bookmarksopen=false,breaklinks=true,pdfborder={0 0 0},backref=false,colorlinks=true]{hyperref}

\begin{document}
\title{Terahertz generation in dual-color laser: \\ continuum electron in strong-field single-atom ionization}

\affiliation{Shanghai Advanced Research Institute, Chinese Academy of Sciences, Shanghai 201210, China}
\affiliation{Center for Terahertz Waves and College of Precision Instrument and Optoelectronics Engineering, Key Laboratory of Education, Tianjin University,  Tianjin 300072, China}
\author{Kaixuan Zhang}
\affiliation{Shanghai Advanced Research Institute, Chinese Academy of Sciences, Shanghai 201210, China}
\affiliation{University of Chinese Academy of Sciences, Beijing 100049, China}
\author{Yizhu Zhang}
 \email{zhangyz@sari.ac.cn}
 \affiliation{Shanghai Advanced Research Institute, Chinese Academy of Sciences, Shanghai 201210, China}
 \affiliation{Center for Terahertz Waves and College of Precision Instrument and Optoelectronics Engineering, Key Laboratory of Education, Tianjin University,  Tianjin 300072, China}
\author{T.-M. Yan}
 \email{yantm@sari.ac.cn}
\affiliation{Shanghai Advanced Research Institute, Chinese Academy of Sciences, Shanghai 201210, China}
\author{Y.H. Jiang}
 \email{jiangyh@sari.ac.cn}
 \affiliation{Shanghai Advanced Research Institute, Chinese Academy of Sciences, Shanghai 201210, China}
 \affiliation{University of Chinese Academy of Sciences, Beijing 100049, China}
\affiliation{School of Physical Science and Technology, Shanghai Tech University, Shanghai 201210, China}

\date{\today}

\begin{abstract}
The terahertz (THz) generation in a dual-color field was investigated experimentally by precisely controlling the relative time delay $\tau$ and polarization $\theta$ of dual-color lasers, where the accompanying third-harmonic generation (THG) is employed to determine $\tau$ up to the sub-wavelength accuracy. An anticorrelation of the yields between THz and THG with $\tau$ was displayed. Compared with the theoretical simulations, the experimental results reveal that the continuum-continuum transitions of the released electron after single-atom ionization are the dominating mechanism for THz generation, which is opposite to the high-harmonic generation (HHG) based on the recollision mechanism, or the radiation induced by perturbative bound-bound transitions. 
\end{abstract}

\pacs{Valid PACS appear here}
\maketitle


The enhanced terahertz wave generation (TWG) using dual-color femtosecond pulse, typically by focusing 800 nm ($\omega$) and 400 nm ($2\omega$) beam into gas-phase medium, allows for convenient and efficient access to moderately strong ultra-broadband terahertz field \cite{Cook2000,Kim2008,oh2014apl,Yasuo2013apl}. The approach is applicable in a wide range of fields including THz nonlinear optics \cite{Gaal2006PRL}, remote sensing \cite{Liu2010}, photoelectron streaking \cite{Wimmer2014} and ultrafast spectroscopy \cite{Chen2016}. However, the underlying physical mechanism is still controversial. 

There are two sides to reveal the underlying physics: From the single-atom perspective, TWG originates from the interaction between the strong laser field and an isolated atom \cite{Kostin2016,Zhang2012,Zhou2009,Balciunas2015}. From the plasma perspective, TWG is influenced by time-variant plasma density \cite{Kim2007,Kim2008}, the scattering of electronic wave packet with neighboring atoms \cite{Karpowicz2009}, and the propagation effects \cite{PRL2016Andreeva,PRL2016Zhang}. Actually, the similar argument happened in the early days when pursuing the mechanism of HHG. It is shown that, each individual atom emits HHG radiation, and the waves propagate, interfering and scattering with remaining atoms in gas-phase ensemble, to enhance the further emission. However, obviously the collective behavior does not produce HHG if the single atoms do not. An analogous question is whether or not the same conclusion does suit the explanation for TWG.

Massive efforts have been dedicated to explore the origin of TWG in dual-color fields. For a single atom, although numerically solving the time-dependent Schr\"{o}dinger equation (TDSE) provides a full quantum-mechanical description\cite{Zhou2009,Kostin2016}, the lack of a transparent grasp of underlying physics promotes to pursue straightforward interpretations based on appropriate models. One such model as derived from the perturbation theory considers TWG as a four wave mixing (FWM) process \cite{Cook2000,Xie2006,Zhang2009,Markus2004,Bartel2005}, which has been widely used in nonlinear optical spectroscopy\cite{mukamel1999} and crystal nonlinear optics \cite{RobertBoyd2008}. When interpreting TWG from single-atom perspective, the perturbative susceptibility $\chi^{(3)}$ up to the third-order accounts for the resonant bound-bound (B-B) transitions among Rydberg states in atoms or densely-spaced levels in ions. 

Beyond the perturbative treatment, strong field approximation (SFA), the workhorse in the strong field physics, has also been used to model the TWG. The SFA method provides an intuitive but rather precise descriptions of various strong field phenomena, such as the above-threshold ionization and the high-order harmonic generation (HHG). Especially, the radiation mechanism of HHG is clarified by the SFA as the continuum-bound (C-B) transition when the released electron recombines with the ionic core, also known as the three-step model. The continuum-continuum (C-C) transition may also contribute to the radiation \cite{Becker1997}, although it is often less discussed. Investigations show that the C-C transition also contributes to HHG \cite{Kohler2010} and TWG \cite{Zhou2009} in dual-color field. 

From plasma perspective, the non-perturbative classical photocurrent (PC) model takes the collective behavior into account \cite{Kim2007,Kim2008,Kostin2016}. The PC model suggests that the residual accumulation of the plasma-current density modulated by the dual-color field is responsible for TWG. The quantum-mechanically revised PC model was derived by incorporating both the time-variant electron density and the interference between continuum electronic wave packets \cite{Balciunas2015}.

\begin{figure}[t]
\includegraphics[width=\linewidth]{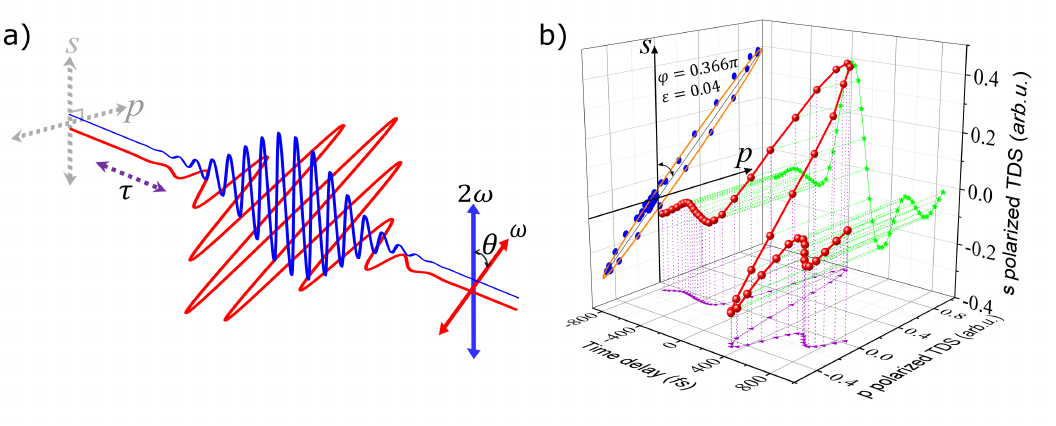}
\caption{\label{fig:experiment-scheme}The schematics of the experiment. (a) The geometry of the measurement. The colinear 400 nm ($2\omega$) and 800 nm ($\omega$) laser pulses focus into the air with the relative time delay $\tau$ and angle $\theta$. (b) The vector of the instantaneous THz electric field (red) is reconstructed with the $s$-polarized (green) and $p$-polarized (magenta) components of the THz electric field. The projection (blue dots) of the vector onto the spatial plane ($E_{s}$-$E_{p}$ plane) is ellipse fitted (orange), from which the predominant polarization $\varphi$ and the ellipticity $\varepsilon$ can be read out.}
\end{figure}

So far, all the above models work to some extent, at least achieving some agreements with currently available experimental results. However, the models are derived from apparently different mechanisms. Questions thus arise. Which model is better for TWG in the dual-color fields? Should the generation process be described perturbatively or not? Is the free electron or the bound electron responsible for TWG? Is TWG similar to the HHG where the three-step recombination mechanism dominates? What is the role from individual atoms and the collective behavior of the ensemble? Above all, is it possible to conduct an appropriate experiment to answer the above questions?

Experiments have been proposed to unravel the mechanism of dual-color-field TWG. One proposal is the investigation of THz features using complex dual-color-field polarization. Comparing TWG in linearly polarized dual-color field, a surprising enhancement of THz signal has been discovered under circularly polarized fields, which is confirmed by TDSE calculations \cite{Meng2016}. Also, the polarization of TWG is modulated with elliptically polarized dual-color fields \cite{Dai2009}. The other approach is the experimental access to the absolute phase delay between the $\omega$-2$\omega$ electric fields, since the maximized TWG appears at $0\pi$ for the perturbative FWM model \cite{Cook2000,Xie2006} and $\pi$/2 for the non-perturbative PC model \cite{Kim2007}. But in practice, the reliable determination of the phase delay is nontrivial. The attempt was made in \cite{li_verification_2012}, experimentally supporting the PC model. The relative phase shift between the THz and HHG radiations was measured, suggesting that the distortion of electronic wave packet induced by the long-range Coulomb potential influences the THz radiation \cite{Zhang2012}. 

In this work, the joint measurement of terahertz time-domian spectroscopy (THz-TDS) and the third-harmonic generation (THG) is performed. Using the THG signal as reference, the relative phase of dual-color fields can be determined up to the sub-wavelength accuracy. The THz electric field was studied by precisely controlling the relative time delay $\tau$ and polarization angle $\theta$ between the $\omega$ and 2$\omega$ fields. The vector of THz electric field, including both the amplitude and polarization, can be reconstructed by polarization-sensitive detection, and the dependence of the THz signal on $\theta$ and $\tau$ is systematically investigated. Compared with theoretical models, the experimental results show a characteristic inconsistency with the perturbative FWM model and nonperturbative C-B model. The results evaluated by C-C and PC models achieve good agreements with the experimental result. We developed the single-atom photocurrent (SPC) model, which well reproduces the major experimental results. Our investigation suggests that the free electron from single-atom ionization dominates the TWG in dual-color field. Manifold implications are inherent in this work. First, the THz radiation is still possible with C-C transition even when the electron in bound states is completely depleted by intense light field. It implies the strong-field THz pulse can be accessible using the extremely intense laser. Also, the TWG, complementary to well-studied HHG and the refreshed investigation on near-threshold harmonic generation, provides a unified view of harmonic radiation throughout the entire spectral range.

Our experimental setup for THz generation in the dual-color fields is shown in Supplementary S1, while the THG are jointly measured. The schematics of the measurement is shown in Fig. \ref{fig:experiment-scheme}(a). The 2$\omega$ wave is alway $s$-polarized, and $\theta$ was varied from 0 to $\pi$/2 by rotating a zero-order dual-wavelength half-wave plate, which acts as a half-wave plate for $\omega$ wave and a full-wave plate for 2$\omega$ wave. Controlling relative polarization with a half-wave plate, instead of rotating BBO crystal, avoids the mixture of the polarization of  \textbf{\^{o}} ray and \textbf{\^{e}} ray in BBO crystal. The $\tau$ was varied by changing the distance from the BBO crystal to the focus, considering that air dispersion gives different refractive indices at $\omega$ and 2$\omega$. Here, the phase delay between $\omega$-2$\omega$ pulses is passively stabilized up to sub-wavelength accuracy. The THz-TDS was detected with the polarization-sensitive electro-optic sampling (EOS), and the instantaneous THz electric field can be reconstructed when both $s$-polarized and $p$-polarized waveforms are obtained. The instantaneous THz electric field provides full information including the peak\textendash peak (P-P) amplitude and phase, polarization and ellipticity. 

The yields of THz and THG signals are simultaneously recorded with varying $\tau$ and $\theta$. The instantaneous vector of THz electric field is reconstructed by $\boldsymbol{E}_{\text{THz}}(t)=E_{\text{THz}}^{s}(t)\boldsymbol{e}_{s}+E_{\text{THz}}^{p}(t)\boldsymbol{e}_{p}$, where $\boldsymbol{e}_{s}$ and $\boldsymbol{e}_{p}$ are unit vectors in $s$ and $p$ polarization (Supplementary S2). Fig. \ref{fig:experiment-scheme}(b) shows the vector of THz electric field at $\theta$=$50$\textdegree $\ $when $\tau=0.33\mathrm{\textrm{ fs}}$. The ellipticity $\varepsilon$ is defined as the component ratio between the minor and major axis of the polarization ellipse (orange line), which is the ellipse-fitting of the projection (blue dots) of the THz electric field onto the $s$-$p$ plane. And the THz orientation $\varphi$, defined as the angle of the major axis with respect to the $s$-direction, can also be obtained. In Fig. \ref{fig:experiment-scheme}(b), since $\varepsilon\sim0.04$, the THz wave is approximately in linear polarization, and the orientation $\varphi=0.366\pi$.

\begin{figure*}[htb]
\includegraphics[width=0.95\linewidth]{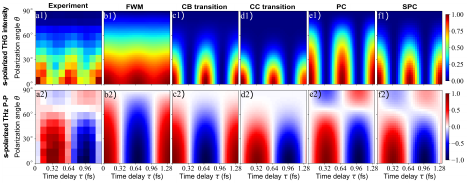}
\caption{\label{fig:joint-measure-delay}The $s$-polarized THz P-P amplitude $E_{\text{THz}}^{s}(\tau,\theta)$ and the THG yield $S_{3\omega}^{s}(\tau,\theta)$ in the joint measurement. (a1) The measured $S_{3\omega}^{s}(\tau,\theta)$ for THG and (a2) $E_{\text{THz}}^{s}(\tau,\theta)$ for THz. $S_{3\omega}^{s}(\tau,\theta)$ and $E_{\text{THz}}^{s}(\tau,\theta)$ with FWM model (b1) and (b2), C-B model (c1) and (c2), C-C model (d1) and (d2), PC model (e1) and (e2), SPC model (f1) and (f2). }
\end{figure*}

The THz P-P amplitude along $s$ polarization, $E_{\text{THz}}^{s}(\tau,\theta)=\pm\mid\max[E_{\text{THz}}^{s}(t)]-\min[E_{\text{THz}}^{s}(t)]\mid$, is shown in Fig. \ref{fig:joint-measure-delay}(a2). While the yield of $s$-polarized THG $S_{3\omega}^{s}(\tau,\theta)$, as jointly measured for comparison, is shown in Fig. \ref{fig:joint-measure-delay}(a1). Both $E_{\text{THz}}^{s}(\tau,\theta)$ and $S_{3\omega}^{s}(\tau,\theta)$ are normalized. We define $E_{\mathrm{THz}}^{s}(\tau,\theta)$ positive when the maximum of THz waveform appears along the positive direction of $s$-axis, and vice versa (Supplementary S2). $S_{3\omega}^{s}(\tau,\theta)$ are modulated with a period of $\sim$0.67 fs, while $E_{\text{THz}}^{s}(\tau,\theta)$ are modulated with $\sim$1.33 fs. The experiment shows that the maximum of THz efficiency appears at $\tau$ where THG yield is minimum. $E_{\text{THz}}^{s}(\tau,\theta=0$\textdegree) and $S_{3\omega}^{s}(\tau,\theta=0$\textdegree) are plotted and highlighted in Supplementary S3, where the TWG and THG show a clearly anti-correlated feature along $\tau$. The similar feature were also observed along the $p$ polarization (Supplementary S4). As shown in Fig. \ref{fig:joint-measure-delay}(a2), when $\theta$ increasing, $E_{\text{THz}}^{s}(\tau,\theta)$ decreases to the minimum at $\theta$=70\textdegree, and oppositely increases when $\theta$ further increasing.

The $E_{\text{THz}}^{s}(\tau,\theta)$ and $S_{3\omega}^{s}(\tau,\theta)$ are calculated with different theoretical models. For the FWM model, the results of $E_{\text{THz}}^{s}(\tau,\theta)$ and $S_{3\omega}^{s}(\tau,\theta)$
along $s$ polarization are shown in Fig. \ref{fig:joint-measure-delay}(b). The THz electric fields can be calculated by $\boldsymbol{E}_{\text{THz}}(t,\tau)\propto\boldsymbol{\chi}^{(3)}(\omega_{\text{THz}},\omega,\omega,-2\omega)\boldsymbol{E}_{\omega}(t)\boldsymbol{E}_{\omega}(t)\boldsymbol{E}_{2\omega}^{*}(t-\tau)$, where $\boldsymbol{E}(t)$ is the electric-field vectors of $\omega$, 2$\omega$ and THz wave, and $\boldsymbol{\chi}^{(3)}(\omega_{\text{THz}},\omega,\omega,-2\omega)$ is the third-order susceptibility tensor for the THz generation (Supplementary S5). 

The signal intensity for the THG in the FWM model reads $S_{3\omega}(\tau)=\int \mathrm{d} t |\boldsymbol{\chi}^{(3)}(3\omega,\omega,\omega,\omega)\boldsymbol{E}_{\omega}^{3}(t)+ \boldsymbol{\chi}^{\mathrm{(3)}}(3\omega,2\omega,2\omega,-\omega)\boldsymbol{E}_{2\omega}^{2}(t-\tau)\boldsymbol{E}_{\omega}^{*}(t)|^{2}$, where $\boldsymbol{\chi}^{\mathrm{(3)}}(3\omega,\omega,\omega,\omega)$ and $\boldsymbol{\chi}^{\mathrm{(3)}}(3\omega,2\omega,2\omega,-\omega)$ are the third-order susceptibility tensors for THG via the sum-frequency ($3\omega=\omega+\omega+\omega$) pathway and the differential-frequency ($3\omega=2\omega+2\omega-\omega$) pathway, respectively. The dependence of THG on $\tau$ had been observed and explained in the previous experiment \cite{Xu2011}. The periodic modulation of $\sim$0.67 fs originates from the interference between different pathways. 

In contrast to the FWM model, the SFA theory treats the field-atom interaction non-perturbatively. The electromagnetic radiation originates from the oscillating dipole moment \cite{Becker1997}
\begin{widetext}
\begin{eqnarray}
\boldsymbol{R}(t) & = & \langle\Psi(t)|\hat{\boldsymbol{r}}|\Psi(t)\rangle\nonumber \\
 & = & \langle\psi_{0}(t)|\hat{\boldsymbol{r}}|\psi_{0}(t)\rangle\nonumber 
- \mathrm{i} \int_{-\infty}^{t}\mathrm{d} t'\langle\psi_{0}(t)|\hat{\boldsymbol{r}}\hat{U}^{V}(t,t')\hat{H}_{I}(t')|\psi_{0}(t')\rangle\nonumber 
+ \mathrm{i} \int_{-\infty}^{t}\mathrm{d} t'\langle\psi_{0}(t')|\hat{H}_{I}(t')\hat{U}^{V}(t',t)\hat{\boldsymbol{r}}|\psi_{0}(t)\rangle\nonumber \\
 &  & + \int_{-\infty}^{t}\int_{-\infty}^{t}\mathrm{d} t'\mathrm{d} t''\langle \psi_{0}(t')|\hat{H}_{I}(t')\hat{U}^{V}(t',t)\hat{\boldsymbol{r}}\hat{U}^{V}(t,t'')\hat{H}_{I}(t'')|\psi_{0}(t'')\rangle,\label{eq:SFA-1}
\end{eqnarray}
\end{widetext}
with $|\Psi(t)\rangle$ the wave function of electron, and $|\psi_{0}(t)\rangle$ the wave function in the initial state. The field-atom interaction is described by $\hat{H}_{I}(t)=-e\hat{\boldsymbol{r}}\cdotp\boldsymbol{E}(t)$ with the laser electric field $\boldsymbol{E}(t)$. It is assumed that the electron is only subject to the intense external light field and the dynamics are governed by the Volkov time-evolution operator $U^{V}(t)$, and the influence of the atomic binding potential is neglected. Eq. (\ref{eq:SFA-1}) consists of four terms. The first term vanishes due to spherically symmetric potential. The second and the third terms, which are mutually complex conjugates, essentially depict the three-step picture including the ionization at $t'$, the free propagation after $t'$, and the recombination at $t$, which results in the HHG due to C-B transition.

The fourth term represents the C-C transition, which describes the harmonic radiation before free electron finally return to the parent ion: the partial wave function of a single electron is ionized respectively by intense field at $t'$ and $t''$ by $\hat{H}_{I}(t')$ and $\hat{H}_{I}(t'')$, then separately propagates in the electric field, represented with operators $\hat{U}^{V}(t',t)$ and $\hat{U}^{V}(t,t'')$, and finally interference at the instant $t$, accompanying with harmonics emission. The results of C-B and C-C processes are shown in Fig. \ref{fig:joint-measure-delay}(c) and (d).

The PC model is essentially applicable for the classical process, which may lack the subtle information from the quantum perspectives. However, the PC model incorporates the collective behavior from the time-variant plasma density. In the PC model, the electron is liberated and subsequently accelerated by strong laser field. The electron density $N(t)$ and free-electron current density $\boldsymbol{j}(t)$ read
\begin{eqnarray}
\frac{\partial N(t)}{\partial t} & = & [N_{g}-N(t)]w(|\boldsymbol{E}(t)|)\nonumber \\
\frac{\partial\boldsymbol{j}(t)}{\partial t} & = & \frac{e^{2}N(t)}{m}\boldsymbol{E}(t),\label{eq:PC}
\end{eqnarray}
with $N_{g}$ the initial density of atmospheric air, $e$ and $m$ the electron charge and mass, $w(|\boldsymbol{E}(t)|)$ the ionization rate evaluated by Keldysh formula in the adiabatic limit \cite{mulser_high_2010}. The THG and THz yields can be read from the transient plasma-current density $\frac{\partial\boldsymbol{j}(t)}{\partial t}$, shown in Fig. \ref{fig:joint-measure-delay}(e). 

All theoretical models depict similar $\tau$-dependence of $S_{3\omega}^{s}(\tau,\theta)$ that the peak occurs at $\tau=0$ fs. Hence, the zero of $\omega$ and 2$\omega$ phase difference can be experimentally determined when $S_{3\omega}^{s}(\tau,\theta)$ is maximized\, and the $\tau$-dependence of TWG can be precisely calibrated according to $S_{3\omega}^{s}(\tau,\theta)$. Using THG as reference, in Fig. \ref{fig:joint-measure-delay}, the $\tau$-dependence of the experimental and theoretical results are compared. In the FWM and C-B models, the THz and THG yields are synchronous with regard to $\tau$, whereas the THz efficiencies is anti-correlated with THG in the C-C and PC models. The PC and C-C models achieve the good agreement with measurements, while the FWM and C-B models predict the opposite.

The failure of the FWM and C-B models implicates that the perturbative susceptibility and the electron re-combination with the parent ion have minor contributions to TWG. The C-C transition shares the similar pattern as in PC model, because in both models the released electron driven by the strong laser is the origin of TWG. The phase shift of the $\tau$-dependent TWG yield is not found in our measurement as the previous studies \cite{Zhang2012}. The main reason is the relatively strong 2$\omega$ intensity, diminishing the influence of the long-range Coulomb potential \cite{Chen2015}. Some rigorous features in Fig. \ref{fig:joint-measure-delay}(a2), that $E_{\text{THz}}^{s}(\tau,\theta)$ has minimum at $\theta=70\text{\textdegree}$, can only be reproduced by the PC model. The fundamental distinctness of the C-C and PC model is that the PC model introduces the collective effect as the plasma density $N(t)$. In order to further clarify the influence of the collective behavior, we develop the single-atom photocurrent (SPC) model:
\begin{equation}
\boldsymbol{E}_{\text{THz}}(t)\propto\frac{e}{m}\int_{-\infty}^{t}w(|\boldsymbol{E}(t')|)\boldsymbol{E}(t') \mathrm{d} t'\label{eq:single-atom PC}
\end{equation}
where the accumulation of $N(t)$ in Eq. (\ref{eq:PC}) is excluded. Here, we assume that $\boldsymbol{E}_{\text{THz}}(t)$ comes from the integral of the acceleration of the released electron at ionization instant $t'$, where the weight is the ionization rate $w(t')$. The result of Eq. (\ref{eq:single-atom PC}) is shown in Fig. \ref{fig:joint-measure-delay}(f). The $E_{\text{THz}}^{s}(\tau=0.33\textrm{ fs},\theta)$ and $E_{\text{THz}}^{s}(\tau,\theta=90\text{\textdegree})$ of the experiment, C-C , PC and SPC models are highlighted and compared in Supplementary S6. The SPC model is in the most close agreement with the experimental result, suggesting the single atom as the fundamental origin of TWG. 

\begin{figure}[t]
\includegraphics[width=\linewidth]{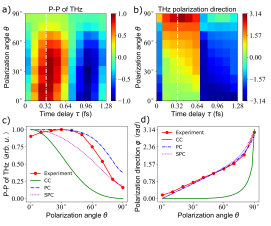}
\caption{\label{fig:joint-measure-angle}(a) $E_{\text{THz}}(\tau,\theta)$ and (b) $\varphi_{\text{THz}}(\tau,\theta)$ varying with $\theta$ and $\tau$. The cross sections of $E_{\text{THz}}(\tau=0.33\textrm{ fs},\theta)$ (c) and $\varphi_{\text{THz}}(\tau=0.33\textrm{ fs},\theta)$ (d), and theoretical predictions with the C-C transition (green line), PC (blue) and SPC model (magenta).
}
\end{figure}

After vector reconstructions of instantaneous THz electric fields, the THz amplitudes $E_{\text{THz}}(\tau,\theta)$ and polarization $\varphi_{\text{THz}}(\tau,\theta)$ varying with $\theta$ and $\tau$ are shown in Fig. \ref{fig:joint-measure-angle}. The $E_{\text{THz}}(\tau,\theta)$ is defined as positive when the maxima of THz electric fields appear at I- and IV- quadrants of $s$-$p$ plane (Supplementary S2). With increasing $\theta$, the efficiency of the THz generations first slightly increases, and suddenly decreases at $\theta\simeq60$\textdegree, shown in Fig. \ref{fig:joint-measure-angle}(a). From the fitting values $\epsilon$ of different polarization ellipses, the THz waves are approximately linearly polarized. $\varphi_{\text{THz}}(\tau,\theta)$ follows the $\theta$ rotation, shown in Fig. \ref{fig:joint-measure-angle}(b). Because the THz signals are relatively weak and the polarization states are quite complex at $\theta=90$\textdegree, the polarization is not convincing at $\theta=90$\textdegree. It is worthwhile that $\varphi$ rotates faster than $\theta$. The region of $\theta$ is from 0\textdegree $\ $to 90\textdegree, yet the $\varphi$ rotates beyond $90$\textdegree.  Another interesting feature is that, at fixed $\theta, \varphi_{\text{THz}}(\tau,\theta)$ continuously changes with $\tau$.

Fig. \ref{fig:joint-measure-angle}(c) and (d) show the cross sections of $E_{\text{THz}}(\tau=0.33\textrm{ fs},\theta)$ and $\varphi_{\text{THz}}(\tau=0.33\textrm{ fs},\theta)$. The C-C transition (green), PC (blue) and SPC model (magenta) are shown for comparison. The C-C model cannot reproduce the data, since the $p$-polarized THz yield predicted by the SFA model is much lower than the experimental result. The discrepancy between PC and SPC model is the accumulation of $N(t)$ in Eq. (\ref{eq:PC}). Fig. \ref{fig:joint-measure-angle}(c) shows that the collective behavior may be taken into account when investigating the total THz electric-field vectors. 

Some characteristic features in Fig. \ref{fig:joint-measure-angle}, e.g., the maximum of the THz yields at $\theta\simeq30$\textdegree, cannot be reproduced by any of above models. Moreover, as shown in Fig. \ref{fig:joint-measure-angle}(b), for the fixed $\theta$, $\varphi_{\text{THz}}(\tau,\theta)$continuously changes versus $\tau$. These features are beyond the prediction of all models. The possible reason is that the SFA model alone cannot precisely predict the properties of low-order harmonics, especially in complex polarization states \cite{Burnett1995,Soifer2010PRL}. The influence of Coulomb potential on the electron wave packet may account for the discrepancy \cite{Zhang2012}. Another possible reason is some more complicated collective effects, for instance, the phase matching in plasma, which is not involved in our analysis. These leave open questions for further investigations.

In this work, a joint measurement of THz electric fields and THG intensity is performed by precisely varying the relative time delay $\tau$ and polarization $\theta$ of $\omega$-2$\omega$ femtosecond pulses, which allows the relative phase delay of dual-color fields to be calibrated up to the sub-wavelength accuracy. Compared with theoretical models, the results present an anticorrelation between THz and THG yields with $\tau$, showing an unequivocal deviation from the perturbative FWM model. The continuum-continnum transition plays a major role, while the recollision process makes no significant contribution to TWG. By comparing single-atom and collective behaviors, the free electron released from single-atom photoionization is considered to be the major origin of the TWG. 

Our conclusion agrees with some existent experimental knowledges of TWG in the dual-color scheme. In C-C transition, continuum electronic wave packet does not need to recombine with the remaining wave packet in the ground state, thus the depletion of a strong field does not hinder THz generation as C-B recombination in HHG, explaining that the focusing power density is not strictly concerned for THz generation as that of HHG. Moreover, it is well known that the continuum electron wave packet cannot recombine with the ground states of other atoms, since there is no phase relationship between electronic wave packets from different atoms. The C-C transition occurs between the continuum states of electron, thus continuum electron wave packets from neighbor atoms probably interfere to emit THz pulse \cite{Karpowicz2009}.

The study was supported by National Natural Science Foundation of China (NSFC) (11420101003, 61675213, 11604347, 91636105), Shanghai Sailing Program (16YF1412600).

\bibliography{THz_single_atom}

\end{document}